# Feedforward Control of DGs for a Self-healing Microgrid

Young-Jin Kim

*Abstract*—**Network reconfiguration (NR) has recently received significant attention due to its potential to improve grid resilience by realizing self-healing microgrids (MGs). This paper proposes a new strategy for the real-time frequency regulation of a reconfigurable MG, wherein the feedforward control of synchronous and inverter-interfaced distributed generators (DGs) is achieved in coordination with the operations of sectionalizing and tie switches (SWs). This enables DGs to compensate more quickly, and preemptively, for a forthcoming variation in load demand due to NR-aided restoration. An analytical dynamic model of a reconfigurable MG is developed to analyze the MG frequency response to NR and hence determine the desired dynamics of the feedforward controllers, with the integration of feedback loops for inertial response emulation and primary and secondary frequency control. A small-signal analysis is conducted to analyze the contribution of the supplementary feedforward control to the MG frequency regulation. Simulation case studies of NR-aided load restoration are also performed. The results of the small-signal analysis and case studies confirm that the proposed strategy is effective for improving the MG frequency regulation under various conditions of load demand, model parameter errors, and communication time delays.**

## I. Introduction

THE frequency and duration of widespread power outages have increased in recent years due to rapid growth in the rate of occurrence of extreme weather events, such as storms and hurricanes. In the United States, weather-related power outages affecting at least 50,000 customers were observed less than 10 times in 1993, but more than 120 times in 2011 [1]. Approximately 90% of the weather-related outages occurred at the distribution level [2]. This highlights the importance of improving resilience (i.e., the ability to prepare for and adapt to changes in grid conditions and recover de-energized loads rapidly) for realization of smart distribution systems [3], [4]. Network reconfiguration (NR) has received significant attention due to its potential to improve the resilience by converting conventional distribution grids into self-healing grids. When power outages occur in a self-healing grid, NR serves to change the topological structure of the grid via the on-off operations of switches (SWs), so that distributed generators (DGs) can supply power to de-energized loads.

In most previous studies on NR (e.g., [5]–[8]), the operations of SWs and DGs were scheduled, for example to maximize the amount of restored load demand and minimize the time taken for restoration. However, the scheduling was carried out considering only the steady-state operation of distribution grids and microgrids (MGs), given the constant or hourly-sampled load demand forecasts during the time periods before and after NR. For the scheduling, DGs were simply treated as point sources (i.e., PQ or PV nodes); their dynamic responses to NR-aided load restoration were not considered. In practice, the SW operations and the corresponding abrupt variations in load demand can cause severe fluctuations in grid frequency and node voltages, given the low moment of inertia of DGs. This can lead to multiple DG tripping and grid voltage collapse. Therefore, further studies on the dynamic effects of NR-aided load restoration are still required.

In [9]–[12], the optimal scheduling of NR was performed considering the dynamic responses of DGs. In [9] and [10], frequency response rate (FRR) models were used to predict the frequency nadirs due to load restorations. The FRR models were then integrated into the constraints of the optimization problems for NR scheduling, so that the maximum frequency deviation did not exceed an acceptable range (e.g., ± 0.5 Hz). Synchronous generators (SGs) were mainly considered as the DGs, although they were modeled in a somewhat simplified form. In [11] and [12], a trial-and-error approach was adopted to consider the dynamic responses of DGs to NR. Numerical simulations were repeatedly conducted to pre-select a set of loads that could be restored while satisfying the constraints on transient variations in the grid frequency and node voltages. However, the load restoration was achieved using only SGs, rather than inverter-interfaced generators (IGs) with faster dynamic responses. Moreover, in [9]–[12], the strategies to control the power outputs of DGs remained unchanged, implying that the frequency and voltage regulation during NR-aided load restoration can be further improved.

A few recent studies (e.g., [13] and [14]) have focused on the real-time control of DGs, in coordination with SW operations, to enhance frequency regulation (FR) during load restoration. Briefly, the feedback loops for the secondary frequency control (SFC) of SGs in [13], and of IGs in [14], were adapted, so that the reference power outputs of the DGs were determined based on information received from SWs via communications systems, such as the on-off statuses and terminal voltages of SWs and the power flowing through SWs. However, the feedback control loops are activated only after the load demand variation due to NR significantly affects the MG frequency. This has motivated the development of new frequency control strategies, wherein DGs can preemptively compensate for the load variation, given that in general, NR is initiated in a controlled manner. In power grids, the preemptive control has become increasingly feasible, as modeling and parameter estimation techniques have continued to be developed.

To develop such strategies, the research gap in the existing literature needs to be considered particularly with regard to NR model accuracy and the applicability of DG control. Specifically, in [5]–[12], NR was modeled simply as the variation in the net load demand (i.e., the amount of load demand to be restored or shed), rather than as a discrete change in the network topology. This can compromise the accuracy of analyses of the dynamic responses of DGs to SW operations and, consequently, the practical performance of NR-aided load restoration. Furthermore, in [13] and [14], the power outputs of DGs were adjusted in advance of NR to maintain the difference between both terminal voltages of each SW at zero; otherwise, large frequency deviations are likely to occur in the transient state. However, the DG control can be achieved only when the feeders on both sides of the SW are in operation. It is not applicable when the feeder on either side is de-energized (i.e., when the terminal voltage is zero).

This paper proposes a new strategy for a reconfigurable MG, wherein the feedforward control of the power outputs of SGs and IGs is achieved in coordination with the on-off operations of SWs to reduce the frequency deviation due to NR-aided load restoration. The feedforward controllers (FFCs) are developed using an analytical dynamic model of the MG. The FFCs are integrated with the feedback loops for primary frequency control (PFC) and SFC to ensure the power sharing among the DGs and the zero frequency deviation in the steady state, respectively. Small-signal analysis is conducted to evaluate the contribution of the proposed strategy to the real-time FR. Simulation case studies are also carried out to confirm the effectiveness and robustness of the proposed strategy.

The main contributions of this paper are summarized below:

• To the best of our knowledge, this is the first study to report feedforward control of SGs and IGs in coordination with SW operations to improve the real-time FR in a reconfigurable MG during NR-aided load restoration.

• The FFCs significantly reduce the frequency deviation due to NR, which cannot be achieved by adjusting the gains for PFC and SFC. The FFCs can be readily integrated with the feedback loops for PFC and SFC, requiring only minor modification of existing FR systems.

• The analytical dynamic model of a reconfigurable MG, discussed in the authors' previous work [15], has been further extended to design the FFCs considering the dynamic responses of DGs, voltage-dependent loads, and distribution line losses to SW operations.

• Comparative case studies of the proposed and conventional FR strategies are conducted under various conditions with respect to load demand, errors in the estimates of DG model parameters, and the communication time delays of the FFCs.

## II. Preemptive Control of DGs in a Reconfigurable MG

### A. Framework

Fig. 1 shows a schematic diagram of the proposed strategy, wherein the supplementary FFCs enable the SGs and IGs to preemptively compensate for a forthcoming variation in load demand due to NR. This aims at improving the real-time FR in a reconfigurable MG, compared to the case where only the PFC and SFC are adopted. In this paper, the FFCs have been developed in the form of transfer function using commonly available information on the MG (see Sections II-B and II-C). The coefficients of the transfer functions are updated online based on the current load demand and the locations of target SWs to better reflect the time-varying dynamics of the MG, as in a common multi-controller architecture [16]. The FFCs are activated in response to the NR-initiating signals, each of which varies from zero to one to change the present state (i.e., open or closed) of the corresponding SW. The FFCs are installed in the same locations as the individual DGs; however, the coefficient updates and NR-initiating signal delivery occur only at the moments of SW operation, mitigating the requirement for communications systems. As shown in Fig. 1, the FFCs are integrated with the feedback loops for the PFC and SFC. The FR reference signals for the PFC are produced at the locations where the DGs are connected to the MG. For the SFC, the reference signals are centrally generated and distributed to the DGs, for example, every 2 s [17]. Note that for the IGs, the additional feedback loops for inertial response emulation (IRE) are considered to further exploit their fast dynamics.

### B. Dynamic Model of a Reconfigurable MG

To design the FFCs, we have estimated the dynamic responses of the MG frequency and node voltages to the SW operations and the corresponding variations in the load demand. In this paper, for the estimation, an analytical model of the reconfigurable MG is developed based on the authors' previous work [15], where the NR was considered as a change in MG topology itself, unlike other previous studies where NR was treated as the amount of load to be restored or shed. This enables more accurate estimation of the MG frequency and voltage variations, improving the performance of the FFCs.

Briefly, in [15], the relationship between *dq*-axis node

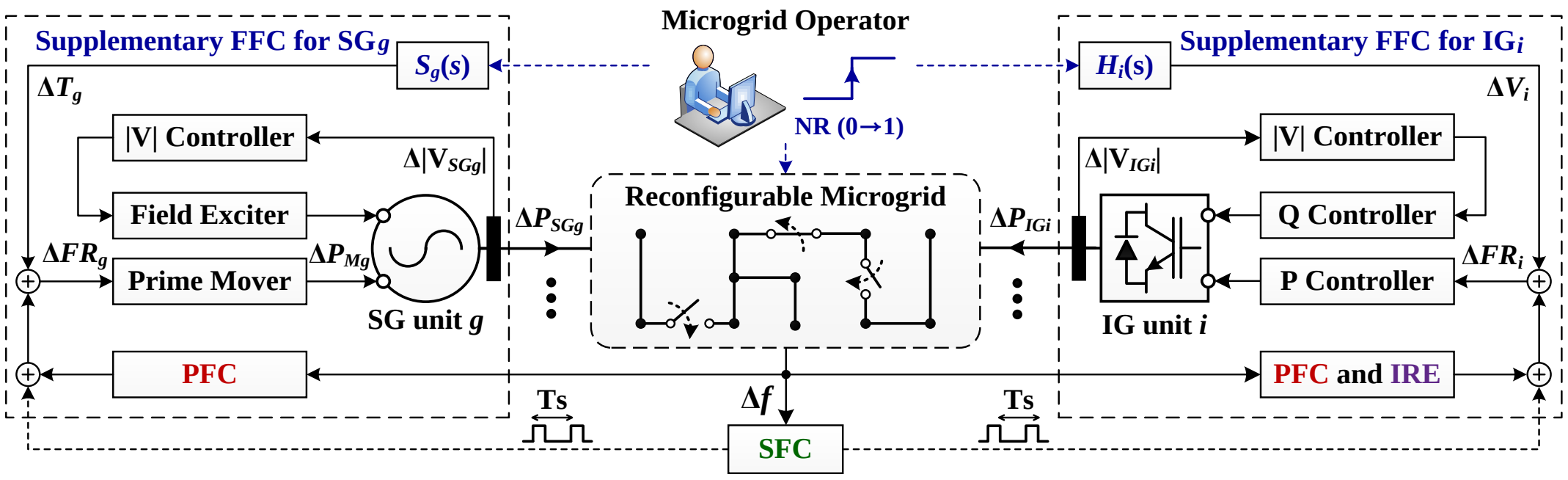

Fig. 1. Schematic diagram of the proposed FR strategy for a reconfigurable MG.

voltages and injection currents at steady state is represented as:

$$\mathbf{I_0} = \mathbf{Y_B}\cdot\mathbf{V_0}. \tag{1}$$

In (1), $\mathbf{Y_B}$ consists of block matrices, where diagonal and off-diagonal blocks for the line between nodes $j$ and $k$ are given by:

$$\mathbf{Y_B}_{jj} = -\sum_{k\neq j}^{N}\mathbf{y}_{jk} \quad \text{and} \quad \mathbf{Y_B}_{jk} = \mathbf{y}_{jk} = \begin{bmatrix} G_{jk} & -B_{jk} \\ B_{jk} & G_{jk} \end{bmatrix}. \tag{2}$$

For simplicity, three-phase (3-*ph*) balanced lines are considered; the analytical modeling of the MG and hence the design of the FFCs can also be applied to the case of unbalanced lines by adopting the 3-*ph* expressions of (1) and (2). After NR is initiated, (1) changes to:

$$\mathbf{I_0} + \mathbf{\Delta I} = \mathbf{Y_A}\cdot(\mathbf{V_0} + \mathbf{\Delta V}). \tag{3}$$

From (1) and (3), $\mathbf{\Delta I}$ is given by:

$$\mathbf{\Delta I} = \mathbf{Y_A}\cdot\mathbf{\Delta V} + (\mathbf{Y_A} - \mathbf{Y_B})\cdot\mathbf{V_0} = \mathbf{Y_A}\cdot\mathbf{\Delta V} + \mathbf{\Delta I_T}, \tag{4}$$

where NR is considered as a discrete variation in the admittance matrix $\mathbf{\Delta Y}$ (i.e., from $\mathbf{Y_B}$ to $\mathbf{Y_A}$), leading to a step variation $\mathbf{\Delta I_T}$ that arises immediately after the MG topology changes. In (4), $\mathbf{\Delta I_T}$ does not decay with time, but drives the MG to another steady state, affecting voltage-dependent loads and line power losses in both transient and steady states.

For $\mathbf{\Delta V}$ in (4), the dynamic responses of the SGs and IGs can be modeled in an aggregated form as:

$$\mathbf{\Delta \dot{X}_{DG}} = \mathbf{A_{DG}}\cdot\mathbf{\Delta X_{DG}} + \mathbf{B_{DG}}\cdot\mathbf{\Delta V}, \tag{5}$$

$$\mathbf{\Delta I_{DG}} = \mathbf{C_{DG}}\cdot\mathbf{\Delta X_{DG}} - \mathbf{D_{DG}}\cdot\mathbf{\Delta V}, \tag{6}$$

where $\mathbf{\Delta X_{DG}} = [\mathbf{\Delta X_{SG}}, \mathbf{\Delta X_{IG}}]$, $\mathbf{\Delta X_{SG}} = [\mathbf{\Delta X_{SG1}}, \cdots, \mathbf{\Delta X_{SG}}_G]$, and $\mathbf{\Delta X_{IG}} = [\mathbf{\Delta X_{IG1}}, \cdots, \mathbf{\Delta X_{IG}}_I]$. In (5) and (6), $\mathbf{A_{DG}}$, $\mathbf{B_{DG}}$, $\mathbf{C_{DG}}$, and $\mathbf{D_{DG}}$ are block diagonal matrices, where the blocks corresponding to $\mathbf{\Delta X_{SG}}_g$ include the model parameters of SG unit $g$ and its prime mover and field exciter [18], as shown in Fig. 1. Similarly, for $\mathbf{\Delta X_{IG}}_i$, the blocks contain the linearized model parameters of IG unit $i$ and its *PQ* controllers [19]. In this paper, a local voltage control scheme has been adopted for simplicity, as in [20] and [21], where the PI controllers of the SGs and IGs are used to maintain their terminal voltages at the values determined under the normal MG condition (i.e., before faults). Moreover, the composite response of voltage-dependent loads can be represented as:

$$\mathbf{\Delta I_L} = \mathbf{D_L}\cdot\mathbf{\Delta V}, \tag{7}$$

where $\mathbf{D_L}$ is a block diagonal matrix with the elements determined based on the ZIP coefficients of the loads [18].

Since $\mathbf{\Delta I} = \mathbf{\Delta I_{DG}} + \mathbf{\Delta I_L}$, the dynamic model of the reconfigurable MG can be established by integrating (4) with (5)–(7). Specifically, $\mathbf{\Delta V}$ can be expressed with respect to $\mathbf{\Delta X_{DG}}$ and $\mathbf{\Delta I_T}$ by substituting (6) and (7) into (4), as:

$$\mathbf{\Delta V} = \mathbf{Z}\cdot(\mathbf{C_{DG}}\cdot\mathbf{\Delta X_{DG}} - \mathbf{\Delta I_T}), \tag{8}$$

where $\mathbf{Z} = (\mathbf{Y_A} + \mathbf{D_{DG}} - \mathbf{D_L})^{-1}$. Using (6) and (8), the total variation in the DG power outputs is given by:

$$\Delta P_{DG} \approx \mathbf{V_0}^T\cdot\mathbf{\Delta I_{DG}} + \mathbf{I_{DG0}}^T\cdot\mathbf{\Delta V},$$

$$= \mathbf{K_X}\cdot\mathbf{\Delta X_{DG}} - \mathbf{K_I}\cdot\mathbf{\Delta I_T}, \tag{9}$$

$$\text{where } \mathbf{K_X} = (\mathbf{V_0}^T\cdot\mathbf{Z}^{-1} - \mathbf{V_0}^T\cdot\mathbf{D_{DG}} + \mathbf{I_{DG0}}^T)\cdot\mathbf{Z}\cdot\mathbf{C_{DG}}, \tag{10}$$

$$\text{and } \mathbf{K_I} = (\mathbf{V_0}^T\cdot\mathbf{D_{DG}} - \mathbf{I_{DG0}}^T)\cdot\mathbf{Z}. \tag{11}$$

Furthermore, using (8), (5) can be equivalently expressed as:

$$\mathbf{\Delta \dot{X}_{DG}} = \mathbf{A_{MG}}\cdot\mathbf{\Delta X_{DG}} + \mathbf{B_{MG}}\cdot\mathbf{\Delta I_T}, \tag{12}$$

where $\mathbf{A_{MG}} = \mathbf{A_{DG}} + \mathbf{B_{DG}}\cdot\mathbf{Z}\cdot\mathbf{C_{DG}}$ and $\mathbf{B_{MG}} = -\mathbf{B_{DG}}\cdot\mathbf{Z}$. In (12), $\mathbf{\Delta X_{DG}}$ includes $\mathbf{\Delta F}$; therefore, $\Delta f$ resulting from NR can be estimated based on the principle of the center of inertia, as:

$$\Delta f = \mathbf{\bar{M}}\cdot\mathbf{\Delta F} = \mathbf{\bar{M}}\cdot\mathbf{S_F}\cdot\mathbf{\Delta X_{DG}}, \text{ where } \mathbf{\bar{M}} = M^{-1}[M_1, \ldots, M_G]. \tag{13}$$

By combining (9), (12), and (13), the dynamic responses of $\Delta P_{DG}$ and $\Delta f$ to NR can be represented in the $s$-domain as:

$$\Delta P_{DG}(s) = [\mathbf{K_X}\cdot(s\mathbf{I} - \mathbf{A_{MG}})^{-1}\cdot\mathbf{B_{MG}} - \mathbf{K_I}]\cdot\mathbf{\Delta I_T}/s = a(s)/s, \tag{14}$$

$$\text{and } \Delta f(s) = [\mathbf{\bar{M}}\cdot\mathbf{S_F}\cdot(s\mathbf{I} - \mathbf{A_{MG}})^{-1}\cdot\mathbf{B_{MG}}]\cdot\mathbf{\Delta I_T}/s = b(s)/s. \tag{15}$$

Consequently, the dynamic variation in the load demand due to NR is estimated using the principle of power conservation, as:

$$\Delta P_L(s) = \Delta P_{DG}(s) + D\cdot\Delta f(s) = (a(s) + D\cdot b(s))/s = p(s)/s, \tag{16}$$

where $1/s$ corresponds to the NR-initiating signal. In (16), $\Delta P_L(s)$ reflects not only the dynamics of the loads to be restored or shed via NR and but also their effects on the variations in the voltage-dependent load demand and line power losses, which cannot be achieved using the conventional MG model [13]. The proposed MG model enables more accurate design of the FFCs, as discussed in Section II-C, so that the DGs can better compensate for $\Delta P_L(s)$ due to NR, improving the regulation of $\Delta f(s)$.

### C. *Feedforward Control of DGs in Response to NR*

Fig. 2 shows a small-signal model of the reconfigurable MG, shown in Fig. 1, where $\Delta f(s)$ can be estimated as:

$$(sM + D)\Delta f(s) = \sum_{g=1}^{G}\Delta P_{Mg}(s) + \sum_{i=1}^{I}\Delta P_{IGi}(s) - \Delta P_L(s). \tag{17}$$

In (17), $\Delta P_{Mg}(s)$ and $\Delta P_{IGi}(s)$ are the responses of SG unit $g$ and IG unit $i$ to the FR reference signals $\Delta FR_g(s)$ and $\Delta FR_i(s)$, respectively, as:

$$\Delta P_{Mg}(s) = t_g(s)\cdot\Delta FR_g(s) \text{ and } \Delta P_{IGi}(s) = v_i(s)\cdot\Delta FR_i(s). \tag{18}$$

Moreover, in (18), $t_g(s)$ and $v_i(s)$ represent the dynamics of the prime mover of SG unit $g$ [22] and of IG unit $i$ with its outer *PQ* and inner current controllers [23], respectively, as:

$$t_g(s) = \frac{sT_{3g}+1}{s^2T_{1g}T_{2g}+sT_{1g}+1}\cdot\frac{sT_{4g}+1}{(sT_{5g}+1)(sT_{6g}+1)}, \tag{19}$$

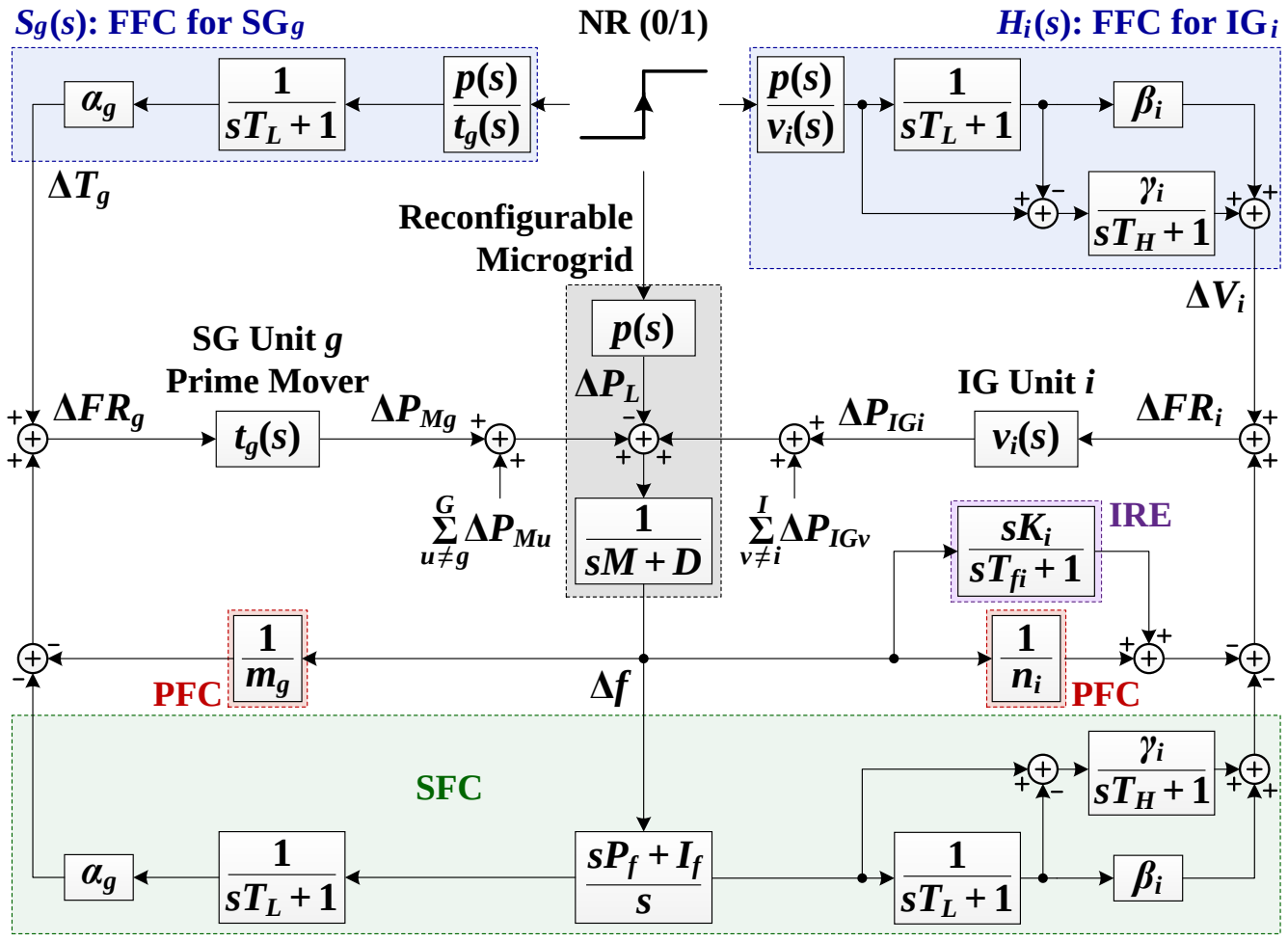

Fig. 2. Small-signal model of the reconfigurable MG with the supplementary FFCs and the feedback loops for the IRE, PFC, and SFC.

$$\text{and} \quad v_i(s) = (sT_{Ei} + 1)^{-1} \ . \tag{20}$$

In the proposed strategy, the FR reference signals consist of two groups; one corresponds to the original feedback loops for the IRE, PFC, and SFC [i.e., $l_g(s)\cdot\Delta f(s)$ and $q_i(s)\cdot\Delta f(s)$] and the other is for the supplementary FFCs [i.e., $\Delta T_g(s)$ and $\Delta V_i(s)$] as:

$$\Delta FR_g = -l_g \Delta f + \Delta T_g \quad \text{and} \quad \Delta FR_i = -q_i \Delta f + \Delta V_i \ , \tag{21}$$

$$\text{where} \quad l_g = \left[ m_g^{-1} + \alpha_g \frac{sP_f + I_f}{s(sT_L + 1)} \right], \tag{22}$$

$$q_i = \left[ n_i^{-1} + \frac{sK_i}{sT_{fi} + 1} + \left( \beta_i + \frac{sT_L \gamma_i}{sT_H + 1} \right) \frac{sP_f + I_f}{s(sT_L + 1)} \right]. \tag{23}$$

Note that for brevity, the notation for the $s$-domain [i.e., ($s$)] was omitted in (21)–(23) and in the equations below. Moreover, the participation factors $\alpha_g$, $\beta_i$, and $\gamma_i$ of the SGs and IGs should satisfy the conditions of $\Sigma_g \alpha_g + \Sigma_i \beta_i = 1$ and $\Sigma_i \gamma_i = 1$.

By substituting (18)–(23) into (17), the closed-loop response of $\Delta f$ to NR can then be estimated as:

$$\Delta f = G_{conv} \cdot \left( \Delta P_L - \sum_{g=1}^{G} t_g \Delta T_g - \sum_{i=1}^{I} v_i \Delta V_i \right), \tag{24}$$

$$\text{where} \quad G_{conv} = -\left( sM + D + \sum_{g=1}^{G} t_g l_g + \sum_{i=1}^{I} v_i q_i \right)^{-1} \ . \tag{25}$$

In (24) and (25), $G_{conv}(s)$ represents the closed-loop response of $\Delta f$ in the conventional case where the DGs respond to only the first group of the FR reference signals: i.e., $l_g(s)\cdot\Delta f(s)$ and $q_i(s)\cdot\Delta f(s)$. It can be seen that $\Delta f(s)$ becomes zero in the ideal case where the second group of the reference signals [i.e., $\Delta T_g(s)$ and $\Delta V_i(s)$] is produced such that the second term on the right-hand side of (24) is set to zero. However, in this paper, $\Delta T_g(s)$ and $\Delta V_i(s)$ have been produced for practical applications, as:

$$\Delta T_g = \frac{\alpha_g \Delta P_L}{(sT_L + 1) t_g} \quad \text{and} \quad \Delta V_i = \left( \beta_i + \frac{sT_L \gamma_i}{sT_H + 1} \right) \frac{\Delta P_L}{(sT_L + 1) v_i} \ . \tag{26}$$

In (26), the low-pass filter with the time constant of $T_L$ enables the SGs to compensate for the low-frequency components of $\Delta P_L(s)$. The IGs compensate for both the low- and high-frequency components of $\Delta P_L(s)$. For the IGs, an additional low-pass filter with $T_H \ll T_L$ is used to prevent the derivative term $sT_L\gamma_i$ from amplifying the high-frequency components of $\Delta P_L(s)$, and thus from inducing excessive operations of the IGs. Moreover, given (16), (26) can be equivalently expressed as:

$$\Delta T_g(s) = S_g(s)/s \quad \text{and} \quad \Delta V_i(s) = H_i(s)/s \ , \tag{27}$$

$$\text{where} \quad S_g(s) = \frac{\alpha_g p(s)}{(sT_L + 1) t_g(s)} \ , \tag{28}$$

$$\text{and} \quad H_i(s) = \left( \beta_i + \frac{sT_L \gamma_i}{sT_H + 1} \right) \frac{p(s)}{(sT_L + 1) v_i(s)} \ , \tag{29}$$

where $S_g(s)$ and $H_i(s)$ represent the transfer functions of the FFCs required for the preemptive control of the DGs.

## III. Small-Signal Analysis

### A. Contributions of the FFCs to Real-time FR

A small-signal analysis of the proposed FR strategy is conducted using the MG model parameters that have been specified in Section IV (see Table II). For brevity, $t_{g=1}(s)$ and $v_{i=1}(s)$ have been assumed to represent the total dynamic responses of the SGs and IGs, respectively (i.e., $G = I = 1$); the analysis results can also be applied to the case of an MG with multiple SGs and IGs. Given the MG condition, the closed-loop response of $\Delta f$ to NR-aided load restoration is represented as:

$$\frac{\Delta f(s)}{\Delta P_L(s)} = G_{conv}(s) \frac{sT_L \cdot sT_H}{(sT_L + 1)(sT_H + 1)} \approx G_{conv}(s) \frac{sT_H}{sT_H + 1} \ , \tag{30}$$

where the approximation is valid for the condition of $T_H \ll T_L$, discussed in Section II-C. As shown in (30), the preemptive control of the DGs via the supplementary FFCs leads to the sig-nificant attenuation of $\Delta P_L(s)$, particularly in the range of $s \ll j1/T_H$, assisting the feedback loops of the IRE, PFC, and SFC to better compensate for $\Delta P_L(s)$. Fig. 3(a) shows a comparison between the Bode plots of $\Delta f(s)/\Delta P_L(s)$ of the proposed and conventional FR strategies, confirming the smaller magnitude of $\Delta f(s)$ for $\Delta P_L(s)$ in the proposed strategy. Moreover, the pole-zero plot, shown in Fig. 3(b), confirms that the proposed strategy ensures the MG frequency stability.

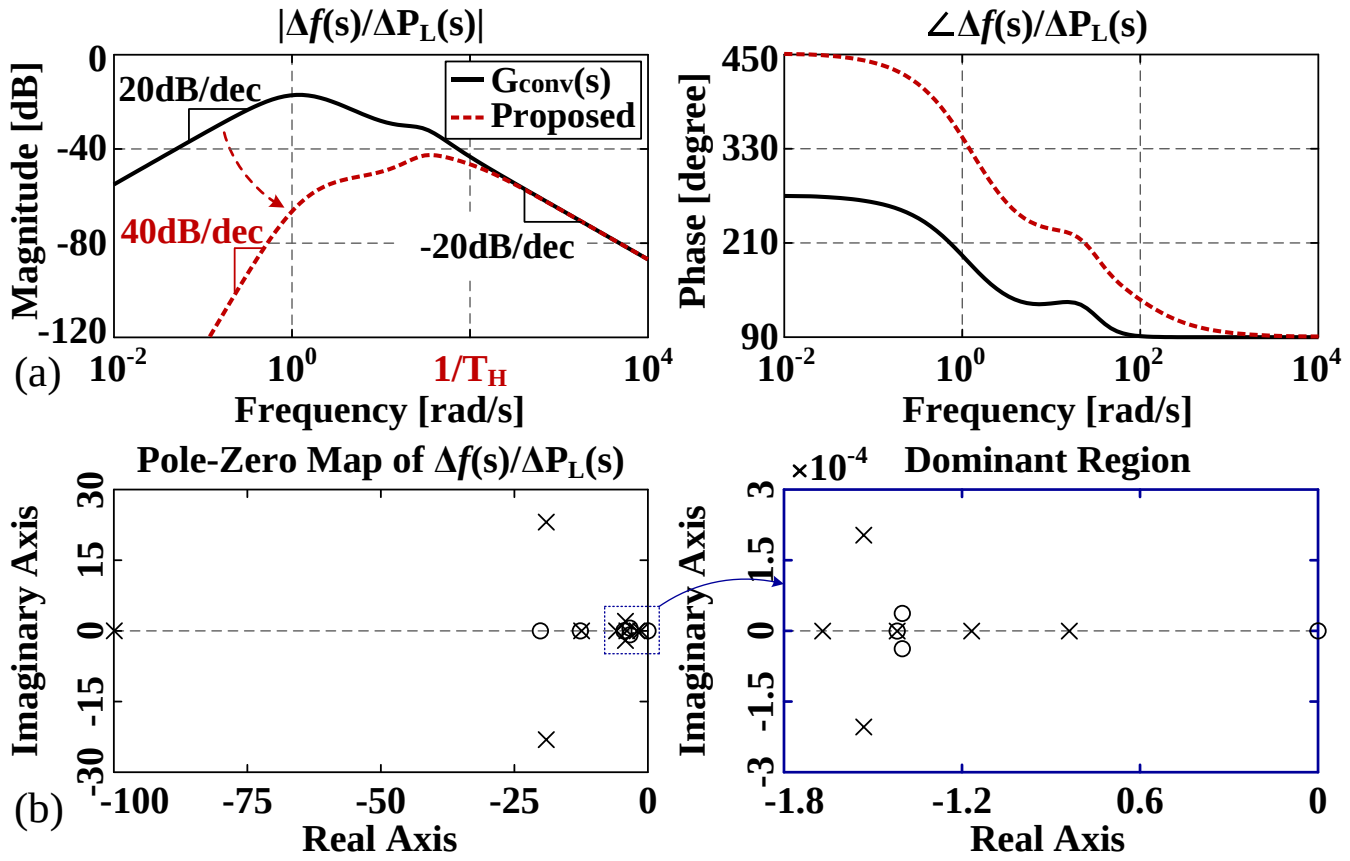

Fig. 3. (a) Bode plots of $\Delta f(s)/\Delta P_L(s)$ for the proposed and conventional FR strategies and (b) pole-zero plot for the proposed FR strategy.

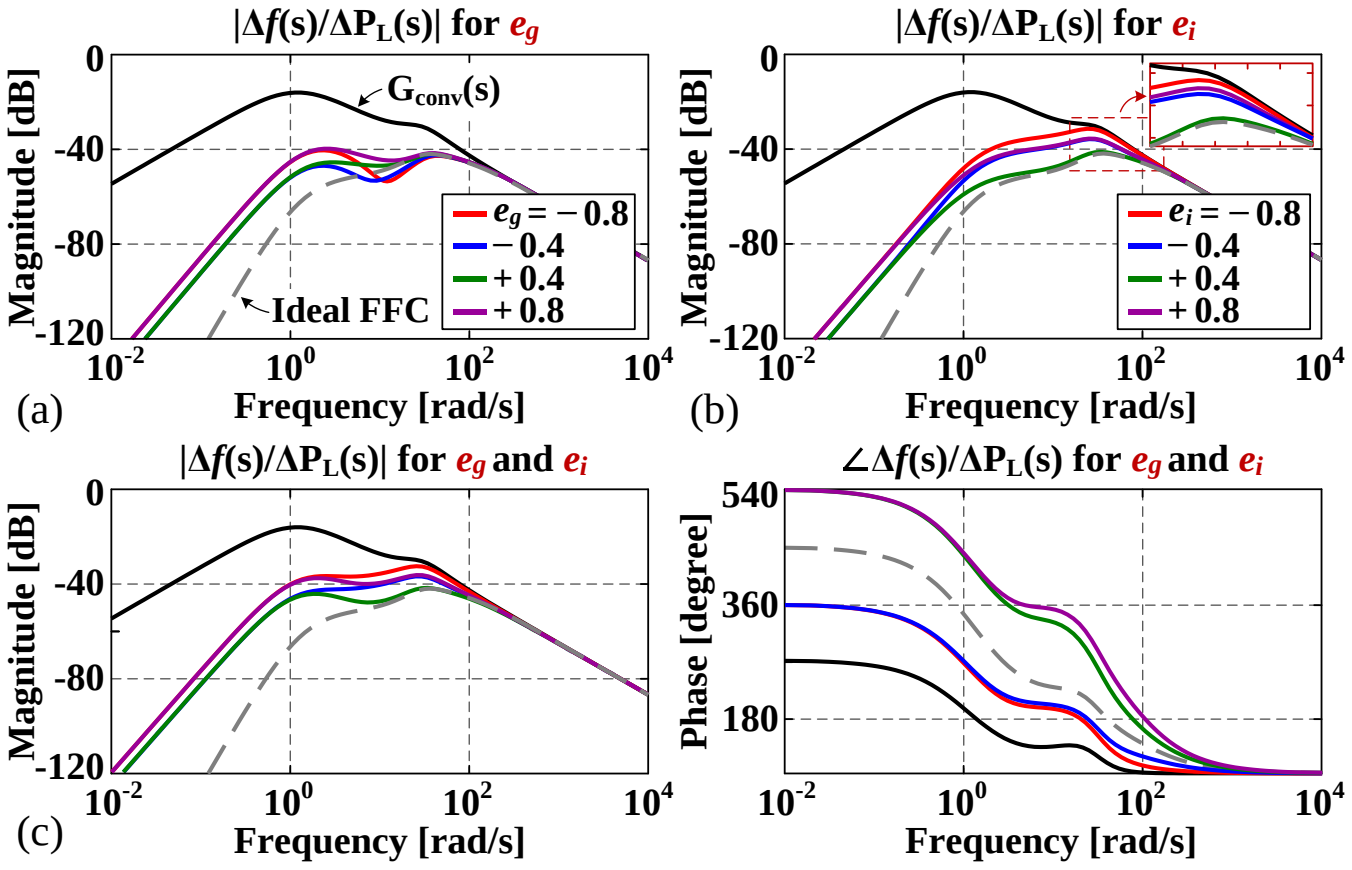

Fig. 4. Bode plots of $\Delta f(s)/\Delta P_L(s)$ for the proposed FR strategy with the errors in the estimates of the DG parameters: (a) $T_{1g}$, (b) $T_{Ei}$, and (c) $T_{1g}$ and $T_{Ei}$.

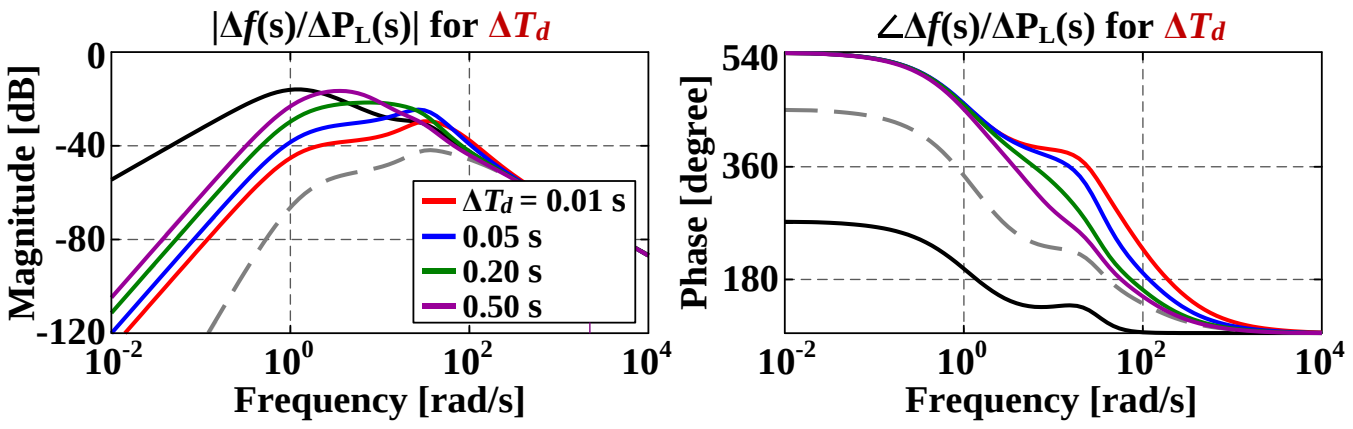

Fig. 5. Bode plots of $\Delta f(s)/\Delta P_L(s)$ for the proposed FR strategy with the communication time delay $\Delta T_d$.

Table I. Comparison of the MG Frequency Responses Between the Proposed and Conventional FR Strategies

| Comparisons | $G_{conv}(s)$ | $G_{prop}(s)$ | | | |
|---|---|---|---|---|---|
| | | $\Delta T_d$ = 0.01 s | 0.05 s | 0.20 s | 0.50 s |
| $\lVert \bullet \rVert_2$ | $5.4 \times 10^{-2}$ | $3.9 \times 10^{-4}$ | $3.0 \times 10^{-3}$ | $1.2 \times 10^{-2}$ | $2.7 \times 10^{-2}$ |
| $\lVert \bullet \rVert_\infty$ | $5.9 \times 10^{-2}$ | $1.4 \times 10^{-3}$ | $1.2 \times 10^{-2}$ | $2.6 \times 10^{-2}$ | $5.9 \times 10^{-2}$ |

### *B. Effects of Model Parameter Errors*

The performance of the proposed strategy is further analyzed particularly considering errors in the estimation of the model parameters of the SGs and IGs. For example, the dynamics of the SG prime mover and the IG unit were assumed to be estimated as $\hat{t}_g(s)$ and $\hat{v}_i(s)$, rather than $t_g(s)$ and $v_i(s)$, in (19) and (20), respectively. This affects the FR reference signals for the FFCs and, consequently, the response of $\Delta f(s)$ to $\Delta P_L(s)$. Specifically, $|\Delta f(s)/\Delta P_L(s)|$ can be approximated for the range of $s \ll j1/T_H$ (i.e., the range where the FFCs mainly contribute to the FR, as shown in Fig. 3(a)) as:

$$|\Delta f(s)/\Delta P_L(s)| = |G_{prop}(s)| \,,$$
$$\approx |G_{conv}(s)| \cdot \left| \frac{\sum_{g=1}^{G} \alpha_g \Delta \bar{t}_g(s)}{(sT_L+1)} + \frac{\sum_{i=1}^{I} (\beta_i + sT_L \gamma_i) \Delta \bar{v}_i(s)}{(sT_L+1)} \right| , \quad (31)$$

$$\text{where} \quad \Delta \bar{t}_g(s) = \left( \frac{t_g(s)}{\hat{t}_g(s)} - 1 \right) \quad \text{and} \quad \Delta \bar{v}_i(s) = \left( \frac{v_i(s)}{\hat{v}_i(s)} - 1 \right). \quad (32)$$

In (31), the second term on the right-hand side is smaller than one, leading to $|G_{prop}(s)| \le |G_{conv}(s)|$, under the conditions as:

$$\left| \sum_{g=1}^{G} \alpha_g \Delta \bar{t}_g(s) + \sum_{i=1}^{I} \beta_i \Delta \bar{v}_i(s) \right| \le 1, \quad \text{for } s \ll j1/T_L \,, \quad (33)$$

$$\text{and} \quad \left| \sum_{i=1}^{I} \gamma_i \Delta \bar{v}_i(s) \right| \le 1, \quad \text{for } s \gg j1/T_L \,. \quad (34)$$

The conditions specify that the magnitude of the uncertainty in the DG response due to the model parameter errors should be smaller than the magnitude of the original DG response itself, which in general is satisfied in practical FFC application [24]. For example, Fig. 4 shows the Bode plots of $\Delta f(s)/\Delta P_L(s)$ [i.e., (31)] for the case where the errors $e_g$ and $e_i$, ranging from –0.8 to 0.8, occur in the estimation of $T_{1g}$ in (19) and $T_{Ei}$ in (20), respectively. In other words, the Bode plots have been obtained using $T_{1g}(1 + e_g)$ and $T_{Ei}(1 + e_i)$ for $-0.8 \le e_g, e_i \le 0.8$, rather than using $T_{1g}$ and $T_{Ei}$, respectively. Note that in most types of SG, $T_{1g}$ is larger than $T_{2g\text{–}6g}$ [25]. For all values of $e_g$ and $e_i$, $|\Delta f(s)/\Delta P_L(s)|$ for the proposed FR strategy still remains lower than that for the conventional strategy, verifying the effectiveness and robustness of the proposed strategy.

### *C. Effects of Communication Time Delays*

Similarly, the proposed FR strategy is analyzed for the case where the FFCs respond to the NR-initiating signals with time delay $\Delta T_d$. The closed-loop response of $\Delta f(s)$ to $\Delta P_L(s)$ is given

$$\text{by:} \quad \Delta f = G_{conv} \cdot \left[ \Delta P_L - \left( \sum_{g=1}^{G} t_g \Delta T_g + \sum_{i=1}^{I} v_i \Delta V_i \right) e^{-s\Delta T_d} \right], \quad (35)$$

where the notation for the $s$-domain is omitted for brevity. Using (26) and (35), $|\Delta f(s)/\Delta P_L(s)|$ can be calculated and then approximated for the range of $s \ll j1/T_H$ as:

$$|\Delta f(s)/\Delta P_L(s)| = |G_{prop}(s)| \approx |G_{conv}(s)| \cdot \left| 1 - e^{-s\Delta T_d} \right| , \quad (36)$$

$$\text{where} \quad \left| 1 - e^{-s\Delta T_d} \right| \approx \left| \frac{\Delta T_d s}{(\Delta T_d^2/12)s^2 + (\Delta T_d/2)s + 1} \right| . \quad (37)$$

In other words, the asynchronous activation of the SWs and FFCs causes $\Delta P_L(s)$ to be less attenuated, compared to the case of the synchronous activation [i.e., (30)]. Moreover, as $\Delta T_d$ increases, the peak of $|\Delta f(s)/\Delta P_L(s)|$ increases and the corresponding value of $s$ is reduced. For example, the maximum value of $\Delta T_d$ (i.e., $\Delta T_{d,max}$) that ensures $|G_{prop}(s)| \le |G_{conv}(s)|$ for all $s$ is calculated as 0.504 s for the MG model parameters provided in Section IV (see Table II). Fig. 5 and Table I show that for $\Delta T_d$ ranging from 0.01s to 0.50 s, the average and peak values of $|\Delta f(s)/\Delta P_L(s)|$ for the proposed FR strategy are maintained below those for the conventional strategy. In [26], it was reported that communication time delays are less than 0.20 s in practical MGs, confirming the effectiveness and wide applicability of the proposed strategy.

## IV. Case Studies and Simulation Results

### *A. Test System and Simulation Conditions*

The proposed FR strategy was tested on the islanded MG, shown in Fig. 6, which was implemented using the IEEE 37-node test feeder with modifications based on [13] and [27]–[29]. The test MG includes tie SWs (TSWs) and sectionalizing SWs (SSWs) that can adaptively change the MG topology in real time. Initially, the TSWs and SSWs were assumed to be open and closed, respectively. The test MG contains four SGs and four IGs, with the total power ratings of 1.68 MVA and 1.24 MVA, respectively. Table II lists their model parameters and the corresponding control gains at the device level; the coefficients used for the feedback loops of the IRE, PFC, and SFC are also provided. Moreover, given the DG power ratings, the total load demand was set to 2.5 + *j*0.8 MVA. For simplicity, the ZIP load coefficients of all nodes were set to 1.4, –2.0, and 1.6 [28], respectively, and 3-*ph* balanced lines were adopted with the impedances determined based on the average value over the three phases for each line configuration [13] and [29]. Note that the proposed strategy can also be applied to an unbalanced MG, as discussed in Section II-B.

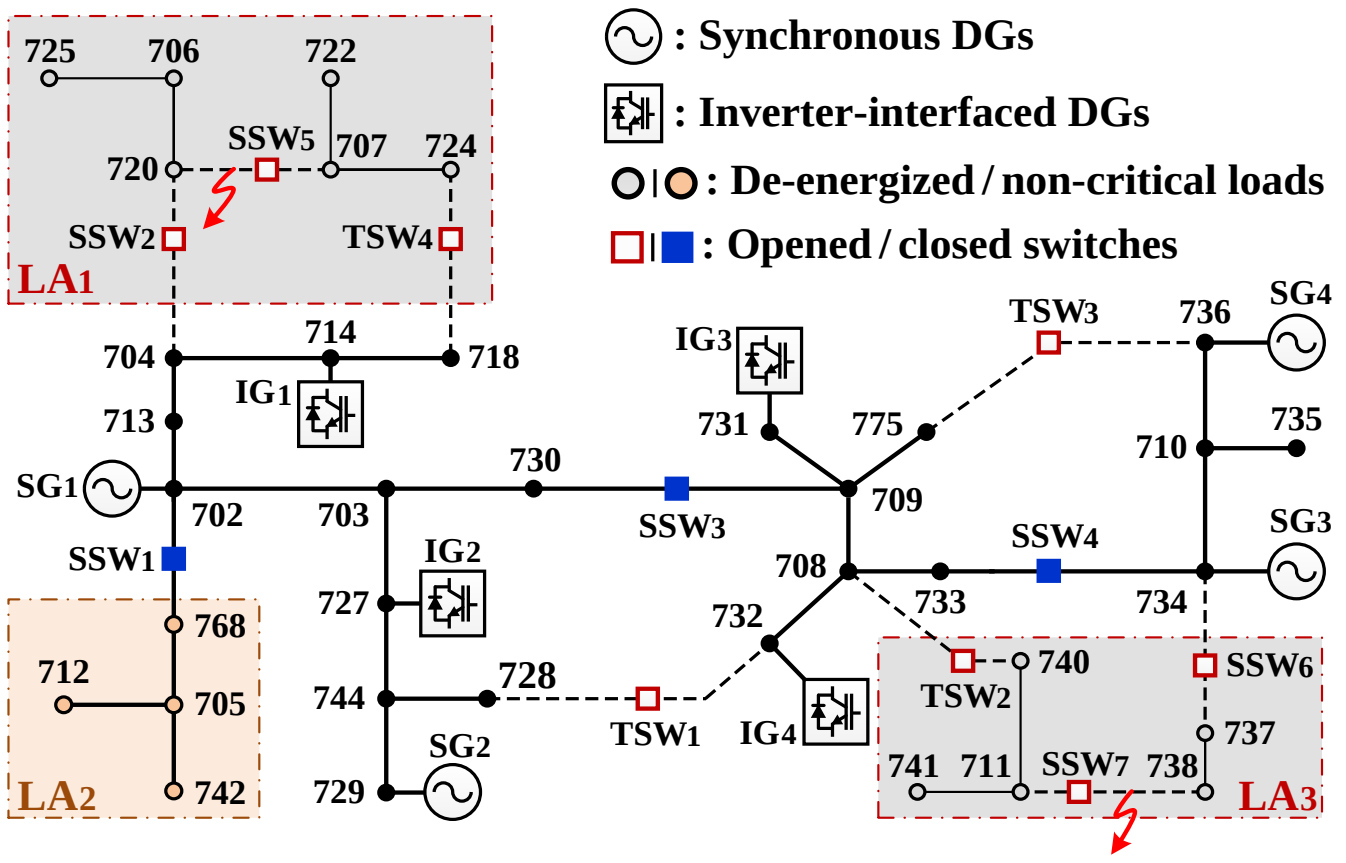

Fig. 6. Single-line diagram of the test MG.

TABLE II. Parameters for the Case Studies

| DGs | Parameters | Values |
|---|---|---|
| SGs (PFC) | $S_{rated}$ [MVA], $V_{ph\text{-}ph}$ [kV$_{rms}$] | 0.42, 4.8 |
| | $M$ [s], $D$ | 2, 0.1 |
| | $X_d$, $X'_d$, $X''_d$ [pu] | 2.24, 0.17, 0.12 |
| | $X_q$, $X'_q$, $X''_q$, $R_s$ [pu] | 1.02, 0.15, 0.13, 0.04 |
| | $T'_{qo}$, $T''_{qo}$, $T'_{do}$, $T''_{do}$ [s] | 4.49, 0.0681, 0.85, 0.034 |
| | $T_1$, $T_2$, $T_3$, $T_4$, $T_5$, $T_6$ [s] | 0.16, 0.03, 0.017, 0.13, 0.08, 0.031 |
| | $T_a$ [s], $K_a$ | 0.02, 200 |
| | $r_l$ [pu/s], $m$ | 0.05, 0.40 |
| IGs (PFC, IRE) | $S_{rated}$ [MVA], $V_{DC}$ [kV] | 0.31, 0.38 |
| | $R_f$ [Ω], $L_f$ [H] | 1.22, 0.05 |
| | $P_i$, $I_i$ | 20, 30 |
| | $T_E$, $T_f$ [s], $n$, $K$ | 0.03, 0.05, 0.10, 5 |
| SFC | $P_f$, $I_f$ | 1, 2 |
| | $T_L$, $T_H$, $T_s$ [s] | 0.58, 0.01, 0.50 |
| | $\alpha_1$, $\alpha_2$, $\alpha_3$, $\alpha_4$ | 0.15, 0.15, 0.15, 0.15 |
| | $\beta_1$, $\beta_2$, $\beta_3$, $\beta_4$ | 0.10, 0.10, 0.10, 0.10 |
| | $\gamma_1$, $\gamma_2$, $\gamma_3$, $\gamma_4$ | 0.30, 0.30, 0.20, 0.20 |

In addition, Table III and Fig. 7 show a test NR scenario, in which the SW operations trigger variation in the MG topology to restore the critical loads in $LA_1$ and $LA_3$. It was assumed that the non-critical loads in $LA_2$ were shed to ensure sufficient reserve capacity of the DGs and then recovered after all the critical loads became energized. In practice, it is common to operate SWs sequentially (i.e., one at a time), rather than simultaneously, to prevent excessive variations in the MG frequency and voltages in the transient state [11]. In this paper, the switching time interval was set to 10 s.

TABLE III. TEST SCENARIO FOR THE CASE STUDIES

| Steps | MG operating status |
|---|---|
| $S_0$. ($t < t_1$) | Two faults occurred at the lines between Nodes 707 and 720 and between Nodes 711 and 738, leading to the opening of $SSW_{2,\,5,\,6,\,7}$. The total output power of the DGs was 0.62 pu under the fault condition. |
| $S_1$. ($t_1 \le t < t_3$) | As shown in Fig. 7(a), $TSW_1$ was closed at $t = t_1$ to reduce the network power loss and hence increase the reserve capacity of the DGs for subsequent load restorations. At $t = t_2$, $SSW_3$ was opened to recover the radial structure of the MG. |
| $S_2$. ($t_3 \le t < t_5$) | The de-energized loads in $LA_1$ were restored by closing $SSW_2$ and $TSW_4$ at $t = t_3$ and $t = t_4$, respectively, as shown in Fig. 7(b). |
| $S_3$. ($t_5 \le t < t_8$) | As shown in Fig. 7(c), $SSW_1$ was opened at $t = t_5$ to shed the non-critical loads in $LA_2$, increasing the reserve capacity of the DGs. Moreover, $TSW_3$ was closed at $t = t_6$ and $SSW_4$ was opened at $t = t_7$ to reduce the network power loss, further increasing the reserve capacity. |
| $S_4$. ($t_8 \le t < t_{10}$) | The de-energized loads in $LA_3$ were restored by closing $TSW_2$ and $SSW_6$ at $t = t_8$ and $t = t_9$, respectively, as shown in Fig. 7(d). |
| $S_5$. ($t \ge t_{10}$) | At $t = t_{10}$, $SSW_1$ was closed to restore the non-critical loads in $LA_1$. The MG operator then terminated the load restoration and examined the faults before returning the MG topology back. |

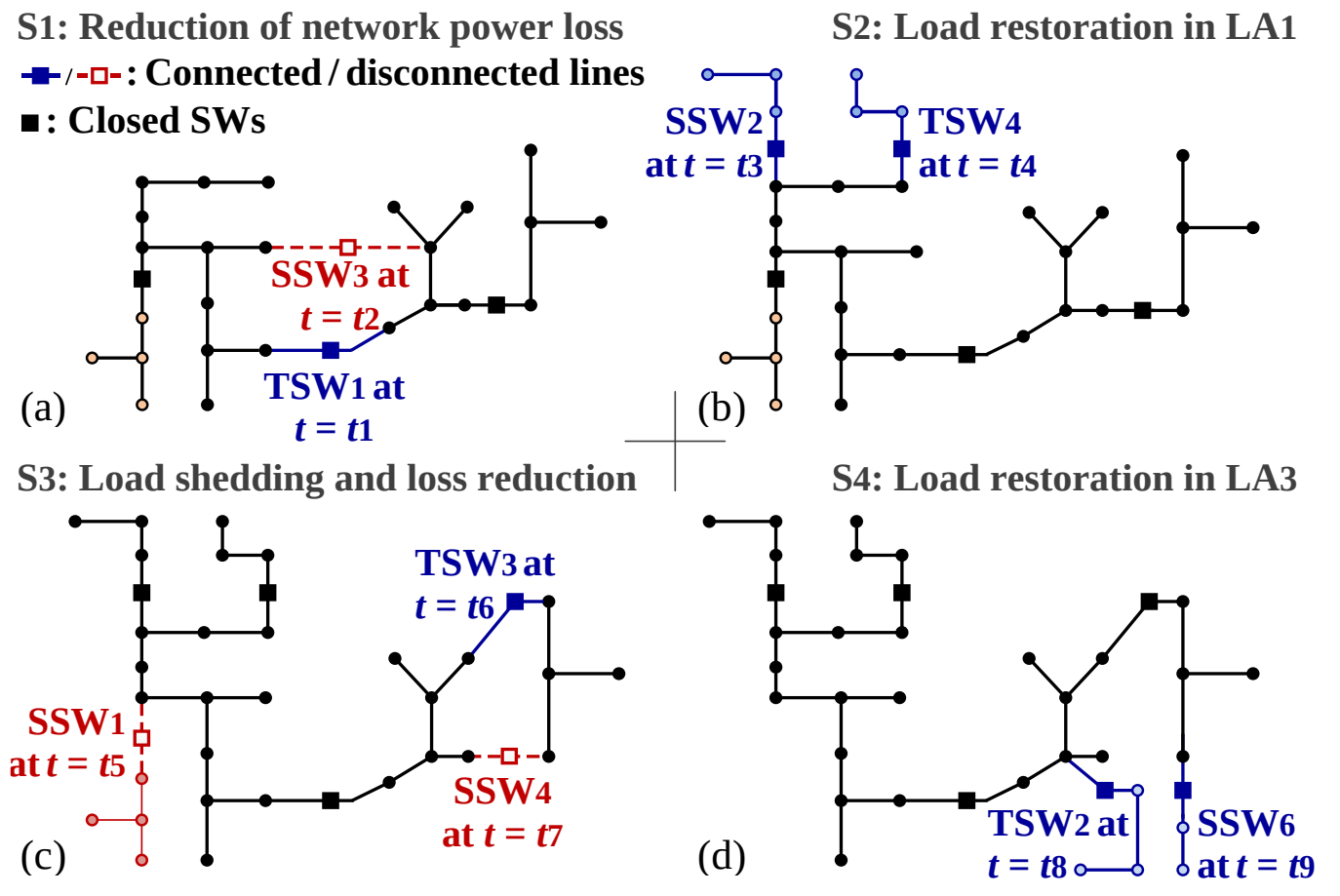

Fig. 7. Variations in the MG topology for the main steps in the test scenario: (a) $S_1$, (b) $S_2$, (c) $S_3$, and (d) $S_4$.

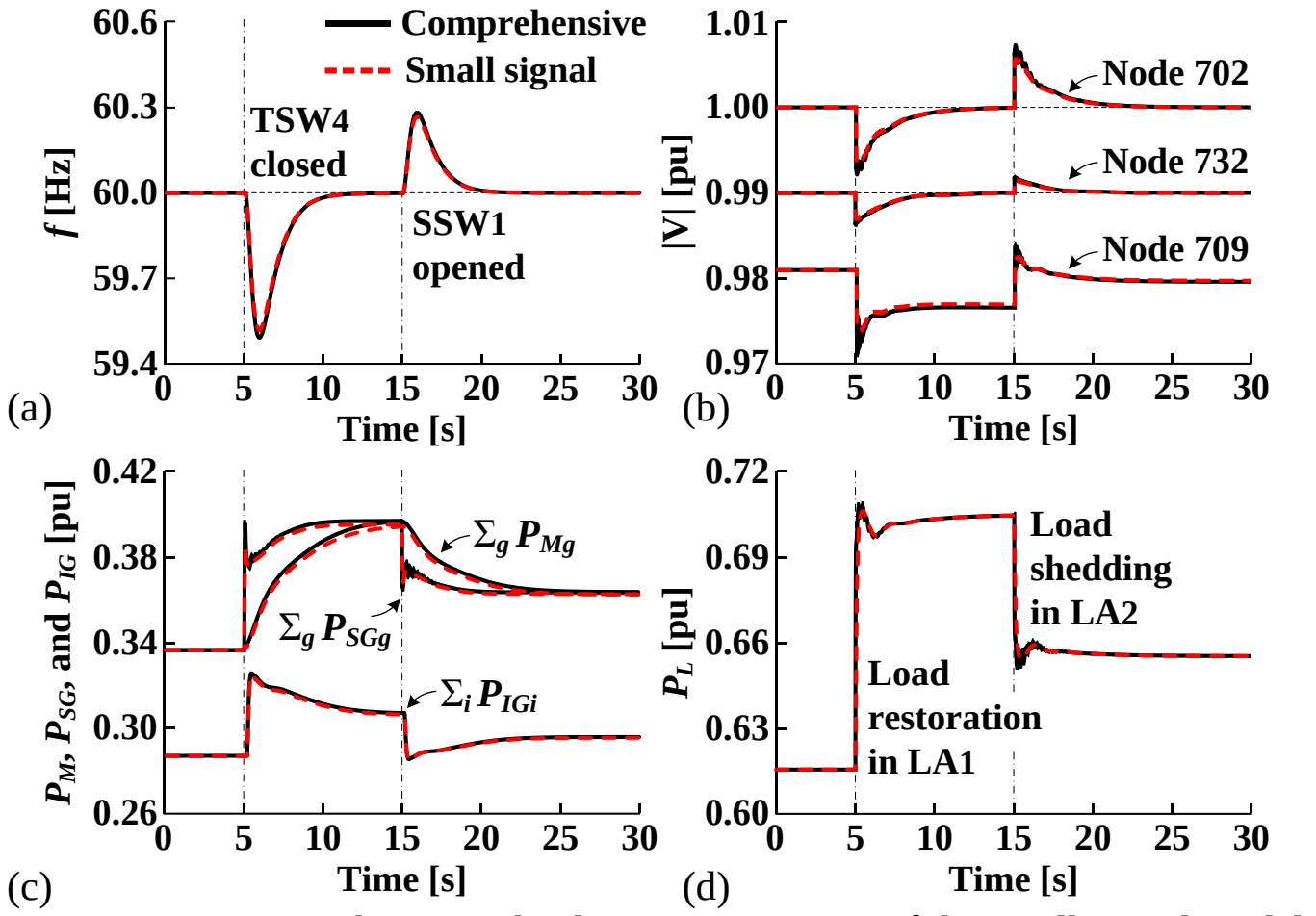

Fig. 8. Comparisons between the dynamic responses of the small-signal model and the comprehensive simulator: (a) $f$, (b) $|\mathbf{V}|$, (c) $P_M$, $P_{SG}$, and $P_{IG}$, and (d) $P_L$.

## *B. Verifying the Small-signal Model of the Reconfigurable MG*

A comprehensive numerical simulation of the test MG was performed using MATLAB/SIMULINK, to provide benchmark profiles of the MG frequency $f$ and node voltage magnitudes $|\mathbf{V}|$. Fig. 8(a) and (b) show the profiles of $f$ and $|\mathbf{V}|$ at arbitrarily selected nodes, respectively, which were obtained using the comprehensive SIMULINK model and the small-signal model, when $TSW_4$ and $SSW_1$ were closed and open at $t$ = 5 s and 15 s, respectively. Note that the SW operations led to the restoration of the critical loads in $LA_1$ and the shedding of the non-critical loads in $LA_2$, respectively. The profiles of $f$ and $|\mathbf{V}|$ for the small-signal model were very similar to those from the comprehensive SIMULINK model in both transient and steady states. This also led to good consistency between the profiles of the DG power outputs and between the net load demand, as shown in Fig. 8(c) and (d), respectively. The comparison results confirm the accuracy of the small-signal model, and hence, that of the case study results presented in Sections IV-C, D, and E.

## *C. Analyzing the Performance of the Proposed FR Strategy*

A comparative analysis of the proposed and conventional FR strategies was conducted for operation of $TSW_4$ and $SSW_1$ operations, as discussed in Section IV-B. Table IV shows the main features of the proposed (Case 1) and conventional (Cases 2–4) strategies. The comparison between Cases 1 and 2 was performed to analyze the effect of the supplementary FFCs on the real-time FR. For a fair comparison, Cases 3 and 4 were also considered to analyze the performance of the proposed strategy in comparison with that of the conventional strategy with increase in the SFC, PFC, and IRE gains; note that for the PFC, the gains are defined as the reciprocals of $m$ and $n$.

Fig. 9(a) and (b) show that the proposed strategy (i.e., Case 1) significantly decreased the maximum deviation of $f$, compared to the conventional strategy (i.e., Cases 2–4), while maintaining the transient variations in $|\mathbf{V}|$ within an acceptable range. Moreover, in Case 1, $f$ was restored back to the nominal value more rapidly, with a smaller overshoot, than in Cases 2–4. This implies that the proposed strategy is effective in reducing the time intervals required for the consecutive operations of SWs, thus facilitating the NR-aided load restoration in practice. The improved performance of the real-time FR was mainly attributed to the supplementary FFCs enabling the DGs to respond faster and preemptively to the upcoming variations in the load demand, resulting from the SW operations. Fig. 9(c) shows that in Case 1, the SGs output more power in the transient state within a shorter period of time than in Cases 2–4. Moreover, Fig. 9(d) shows that the fast response capability of the IGs was better exploited in Case 1, compared to Cases 2–4. Table V shows that in Case 1, $\Delta f_{pk}$ and $\Delta f_{rms}$ were reduced by 86.4% and 89.5%, respectively, compared to those in Case 2. These reductions could not be achieved in Cases 3 and 4. Note that in Cases 3 and 4, further increases in the feedback control gains caused the oscillation of $\Delta f$ and increased the settling time of $\Delta f$ (i.e., $\Delta T_{set}$) [22].

TABLE IV. FEATURES OF THE PROPOSED AND CONVENTIONAL FR STRATEGIES

| FR strategies | | SFC, PFC, and IRE gains |
|---|---|---|
| Proposed | Case 1 | Set as in Table II |
| Conventional | Case 2 | Set as in Table II |
| | Case 3 | $P_f$ = 3 and $I_f$ = 6 for SFC gains |
| | Case 4 | $m$ = 0.30, $n$ = 0.05, and $K$ = 15 for PFC and IRE gains |

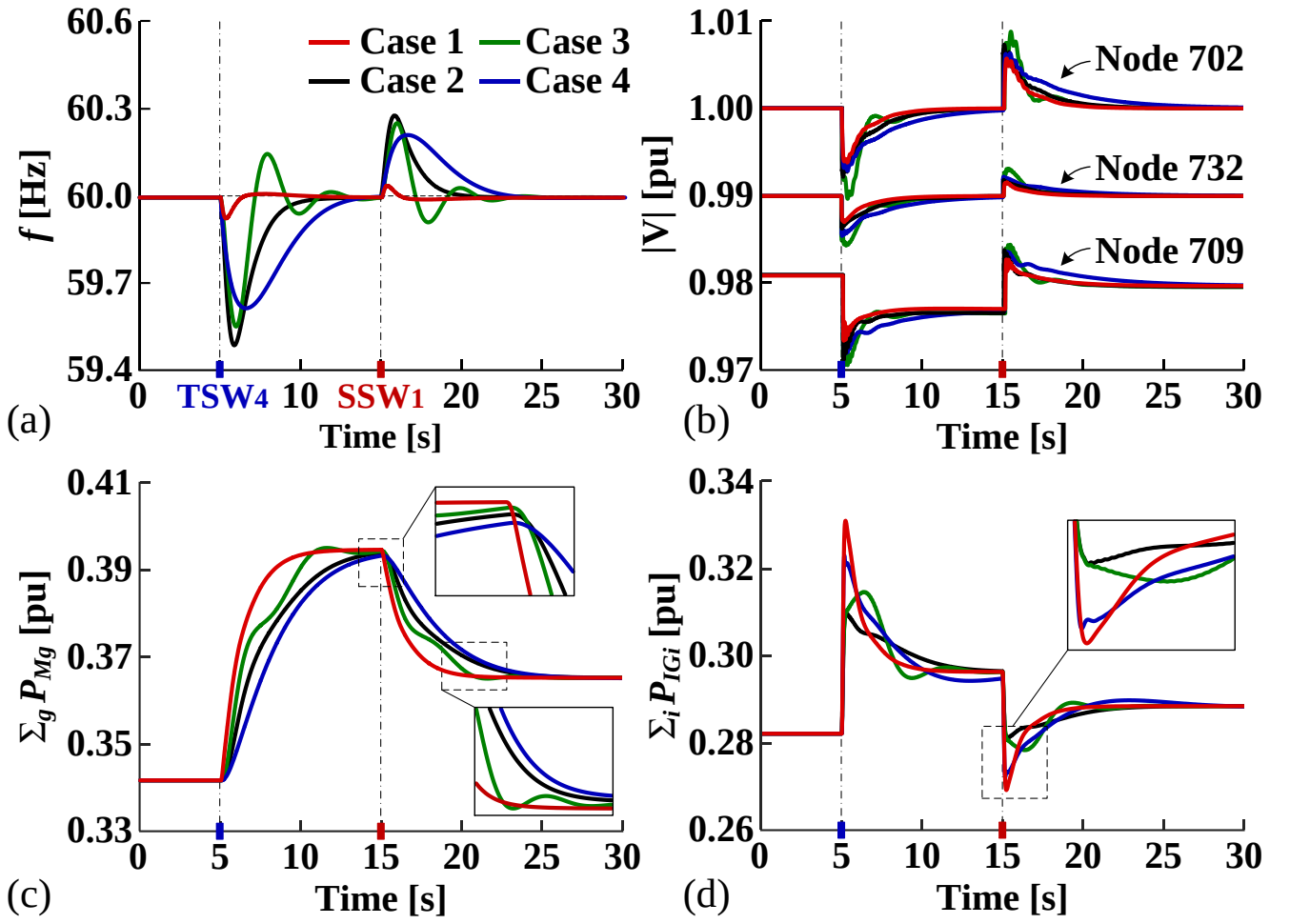

Fig. 9. Comparison between the proposed and conventional FR strategies: (a) $f$, (b) $|\mathbf{V}|$, (c) $\Sigma_g P_{Mg}$, and (e) $\Sigma_i P_{IGi}$.

TABLE V. COMPARISONS FOR THE STEPWISE LOAD VARIATIONS

| Comparison Factors | | (1) Proposed (Case 1) | (2) Conventional Case 2 | Case 3 | Case 4 |
|---|---|---|---|---|---|
| $\Delta f_{pk}$ | [Hz] | 0.11 | 0.81 | 0.69 | 0.60 |
| $\Delta f_{rms}$ | [Hz] | 0.02 | 0.19 | 0.08 | 0.16 |
| $\Delta T_{set}$ | [s] | 2.89 | 5.91 | 6.89 | 8.73 |

### *D. Analyzing the Performance for the Test Scenario*

Additional case studies were performed for the test scenario, discussed in Table III, given the uncertainty in the real-time load demand and renewable power generation. Fig. 10 shows the continuous variations in the MG load demand, reflecting the Reg-D signal of PJM [30]. It also shows the intermittent power outputs of PV arrays [23]. In the case studies, the FFCs were developed on the basis of the base load demand (i.e., 2.5 + $j$0.8 MVA); the difference between the actual and base load demand was reflected as an additional disturbance.

Fig. 11 shows the profiles of $f$, $|\mathbf{V}|$, $\Sigma_g P_{Mg}$, and $\Sigma_i P_{IGi}$ for the steps from $S_0$ to $S_5$ in the test scenario. Fig. 11(a) shows that in Case 1, $\Delta f$ remained significantly lower for all the steps than in

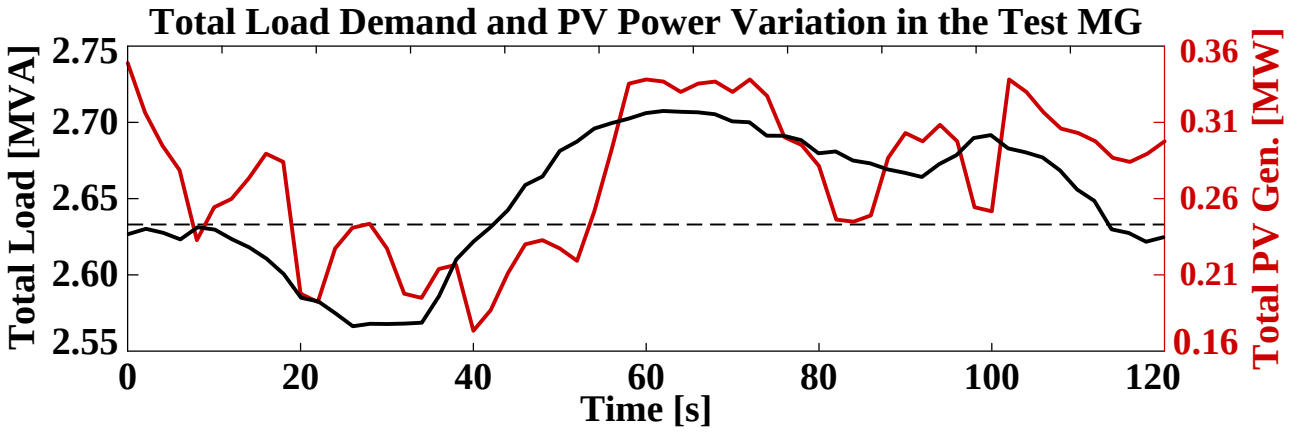

Fig. 10. Continuous variations in the load demand and PV generation.

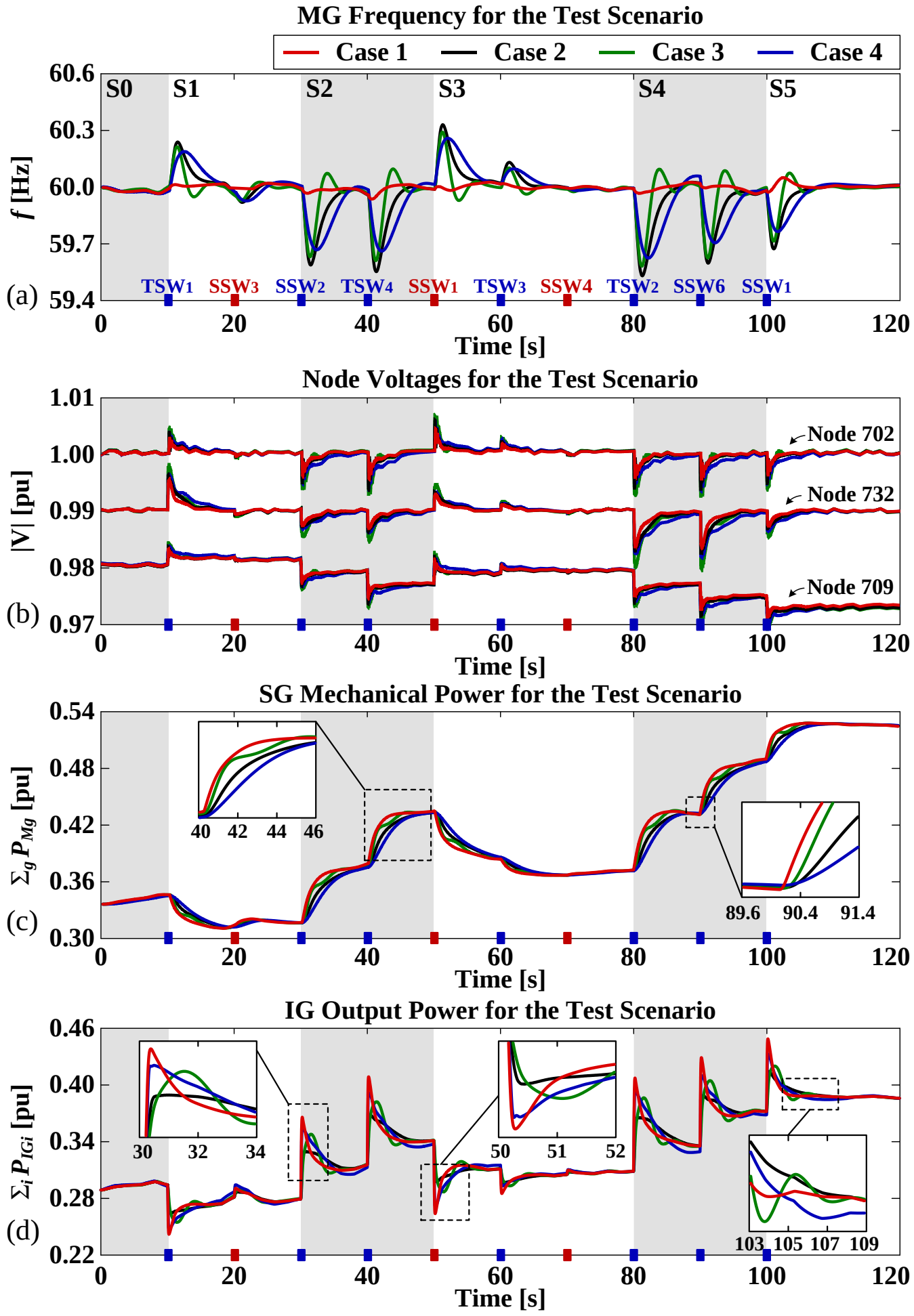

Fig. 11. Comparison between the proposed and conventional FR strategies for the test scenario discussed in Table III: (a) $f$, (b) $|\mathbf{V}|$, (c) $\Sigma_g P_{Mg}$, and (d) $\Sigma_i P_{IGi}$.

TABLE VI. COMPARISONS FOR THE CONTINUOUS LOAD VARIATIONS

| Comparison Factors | | Proposed (Case 1) | Conventional Case 2 | Case 3 | Case 4 |
|---|---|---|---|---|---|
| $\Delta f_{pk}$ | [Hz] | 0.134 | 0.825 | 0.716 | 0.613 |
| $\Delta f_{rms}$ | [Hz] | 0.026 | 0.183 | 0.159 | 0.163 |
| $\Delta P_{M,rms}$ | [pu] | 0.163 | 0.130 | 0.147 | 0.118 |
| $\Delta P_{IG,rms}$ | [pu] | 0.160 | 0.126 | 0.145 | 0.142 |

Cases 2–4. This was mainly because, in the proposed strategy, the supplementary FFCs were able to adjust the DG power outputs preemptively in response to the NR-initiating signals, thus assisting the feedback control loops in compensating for the remains of $\Delta P_L$, as discussed in Sections III-A and IV-C. By contrast, in the conventional strategy, the DG power outputs were controlled only via the feedback loops, and after $\Delta f$ was significantly affected by the abrupt $\Delta P_L$ due to NR. Using the supplementary FFCs, the proposed strategy also enabled faster and larger variation in the DG power outputs in transient state, as shown in Fig. 11(c) and (d). Table VI shows that in Case 1, $\Delta f_{pk}$ and $\Delta f_{rms}$ were 83.8% and 85.8%, respectively, i.e., smaller than in Case 2, whereas $\Delta P_{M,rms}$ and $\Delta P_{IG,rms}$ were only 20.2% and 21.3%, respectively, larger than those in Case 2. This implies that the costs imposed by an increase in the operating stress on the DGs can be compensated for by the savings resulting from the improved FR and the facilitated load restoration.

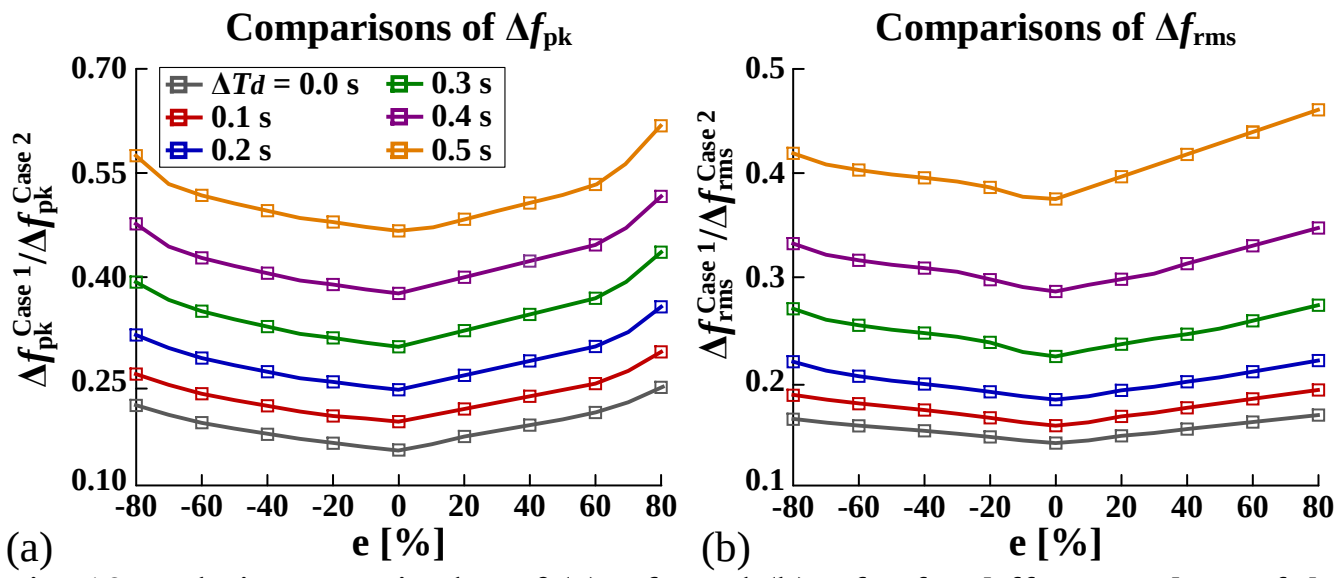

Fig. 12. Relative magnitudes of (a) $\Delta f_{pk}$ and (b) $\Delta f_{rms}$ for different values of the model parameter error $e$ and the communication time delay $\Delta T_d$.

### *E. Sensitivity Analysis*

The case studies, discussed in Section IV-D, were repeated to analyze the sensitivity of the proposed strategy with respect to the model parameter errors and communication time delays. Fig. 12(a) shows the ratio of $\Delta f_{pk}$ for Case 1 to $\Delta f_{pk}$ for Case 2 under the condition where the error $e$ in both estimates of $T_{1g}$ and $T_{Ei}$ varied between –0.8 and 0.8, as discussed in Section III-B. Fig. 12(a) also shows the ratio for the case where the FFCs responded to the NR-initiating signals with time delay $\Delta T_d$, ranging from 0 s to 0.5 s, as discussed in Section III-C. Similarly, Fig. 12(b) shows the ratio of $\Delta f_{rms}$ for Case 1 to $\Delta f_{rms}$ for Case 2 with respect to $e$ and $\Delta T_d$. It can be seen that $\Delta f_{pk}$ and $\Delta f_{rms}$ in Case 1 were remained lower than those in Case 2. This is also the case in the comparisons with Cases 3 and 4, confirming the robustness of the proposed strategy. Moreover, the proposed strategy was less sensitive to the error in the measurement of $f$, because the FFCs were activated by the NR signals.

## V. Conclusions

This paper proposed a new FR strategy for a reconfigurable MG, in which the supplementary FFCs were developed to enable SGs and IGs to respond faster and preemptively to NR-aided load restoration. Using the FFCs, the DGs could take preemptive action to better compensate for a forthcoming variation in the net load demand due to NR. The transfer functions of the FFCs were determined based on the analysis of the MG frequency response to the SW operations. For the analysis, an analytical dynamic model of the MG was developed considering the integration of the FFCs with the feedback loops for the IRE, PFC, and SFC. A small-signal analysis was then conducted, confirming the effectiveness of the proposed strategy in significantly attenuating the low-frequency components of the load demand variation. Moreover, case studies were conducted for a test scenario of the NR-aided load restoration. The proposed strategy decreased the peak-to-peak frequency variations by 86.4% and 83.8%, and the rms variations by 89.5% and 85.8%, for the step and continuous load variations, respectively, compared to the conventional one. The proposed strategy was also proved to be robust against the model parameter errors and communication time delays.